%% file: main.tex
\documentclass[sigconf]{acmart}

\usepackage{enumitem}
\setenumerate[1]{itemsep=0pt,partopsep=0pt,parsep=0pt,topsep=0pt,leftmargin=15pt}
\setitemize[1]{itemsep=0pt,partopsep=0pt,parsep=0pt,topsep=0pt,leftmargin=10pt}
\setdescription{itemsep=0pt,partopsep=0pt,parsep=0pt,topsep=0pt,leftmargin=15pt}

\usepackage{xspace}
\newcommand{\systemName}{\emph{HealthGenie}\xspace}

\input{contents/00_preamable}

\begin{document}

\title[\systemName]{\systemName: Empowering Users with Healthy Dietary Guidance through Knowledge Graph and Large Language Models}


\author{Fan Gao}
\affiliation{%
  \institution{The University of Tokyo}
  \city{Tokyo}
  \country{Japan}
}
\email{fangao0802@g.ecc.u-tokyo.ac.jp}

\author{Xinjie Zhao}
\affiliation{%
  \institution{The University of Tokyo}
  \city{Tokyo}
  \country{Japan}}
\email{xinjie-zhao@g.ecc.u-tokyo.ac.jp}

\author{Ding Xia}
\affiliation{%
  \institution{The University of Tokyo}
  \city{Tokyo}
  \country{Japan}
}
\email{dingxia1995@gmail.com}

\author{Zhongyi Zhou}
\affiliation{%
 \institution{The University of Tokyo}
 \city{Tokyo}
 \country{Japan}}
\email{zhou.zhongyi@tc.u-tokyo.ac.jp}

\author{Rui Yang}
\affiliation{%
  \institution{Duke-NUS Medical School}
  \city{Singapore}
  \country{Singapore}}
\email{yang.rui@duke-nus.edu.sg}

\author{Jinghui Lu}
\affiliation{%
  \institution{Smartor Inc.}
  \city{Shanghai}
  \country{China}}
\email{contact@smartor.ai}

\author{Hang Jiang}
\affiliation{%
  \institution{MIT Media Lab}
  \city{Cambridge}
  \state{Massachusetts}
  \country{USA}}
\email{hjian42@mit.edu}

\author{Chanjun Park}
\affiliation{%
  \institution{Korea University}
  \city{Seoul}
  \country{South Korea}}
\email{bcj1210@naver.com}

\author{Irene Li}
\affiliation{%
 \institution{The University of Tokyo}
 \city{Tokyo}
 \country{Japan}}
\email{bkjl6178@g.ecc.u-tokyo.ac.jp}
\renewcommand{\shortauthors}{Fan Gao, et al.}

\input{section/0_abstract}





\input{contents/00_teaser}
\maketitle

\input{section/1_introduction}
\input{section/2_related_work}

\input{section/3_formative_study}
\input{section/4_usage_senario}
\input{section/5_system_design}
\input{section/6_evaluation}



\input{section/7_results_analysis}

\input{section/8_discussion}

\input{section/9_conclusion}
\begin{acks}
Acknowledgements go here. Delete enclosing begin/end markers if there are no acknowledgements.
\end{acks}

\bibliographystyle{ACM-Reference-Format}
\bibliography{references.bib}

\end{document}

%% file: contents/00_preamable.tex
\usepackage{xcolor}

\usepackage{lipsum} 
\usepackage{graphicx}
\usepackage{multirow}
\usepackage{makecell}
\usepackage{tabularx}

\AtBeginDocument{%
  }

\setcopyright{acmlicensed}
\copyrightyear{2025}
\acmYear{2025}
\acmDOI{XXXXXXX.XXXXXXX}

\acmConference[Conference acronym 'XX]{Make sure to enter the correct
  conference title from your rights confirmation email}{June 03--05,
  2018}{Woodstock, NY}
\acmISBN{978-1-4503-XXXX-X/18/06}

\graphicspath{{./images/}} 

\newif\ifCOMMENTS
\COMMENTStrue
\ifCOMMENTS

\else

\fi

\renewenvironment{quote}[1][0.04\linewidth]
  {\list{}{\leftmargin=#1\rightmargin=#1}\item\relax}{\endlist}

%% file: section/0_abstract.tex
\begin{abstract}

Seeking dietary guidance often requires navigating complex professional knowledge while accommodating individual health conditions. Knowledge Graphs (KGs) offer structured and interpretable nutritional information, whereas Large Language Models (LLMs) naturally facilitate conversational recommendation delivery. In this paper, we present \systemName, an interactive system that combines the strengths of LLMs and KGs to provide personalized dietary recommendations along with hierarchical information visualization for a quick and intuitive overview. Upon receiving a user query, \systemName performs query refinement and retrieves relevant information from a pre-built KG. The system then visualizes and highlights pertinent information, organized by defined categories, while offering detailed, explainable recommendation rationales. Users can further tailor these recommendations by adjusting preferences interactively. Our evaluation, comprising a within-subject comparative experiment and an open-ended discussion, demonstrates that \systemName effectively supports users in obtaining personalized dietary guidance based on their health conditions while reducing interaction effort and cognitive load. These findings highlight the potential of LLM-KG integration in supporting decision-making through explainable and visualized information. We examine the system's usefulness and effectiveness with an N=12 within-subject study and provide design considerations for future systems that integrate conversational LLM and KG.


\end{abstract}

%% file: contents/00_teaser.tex
\begin{teaserfigure}
    \centering
    \includegraphics[width=0.99\linewidth]{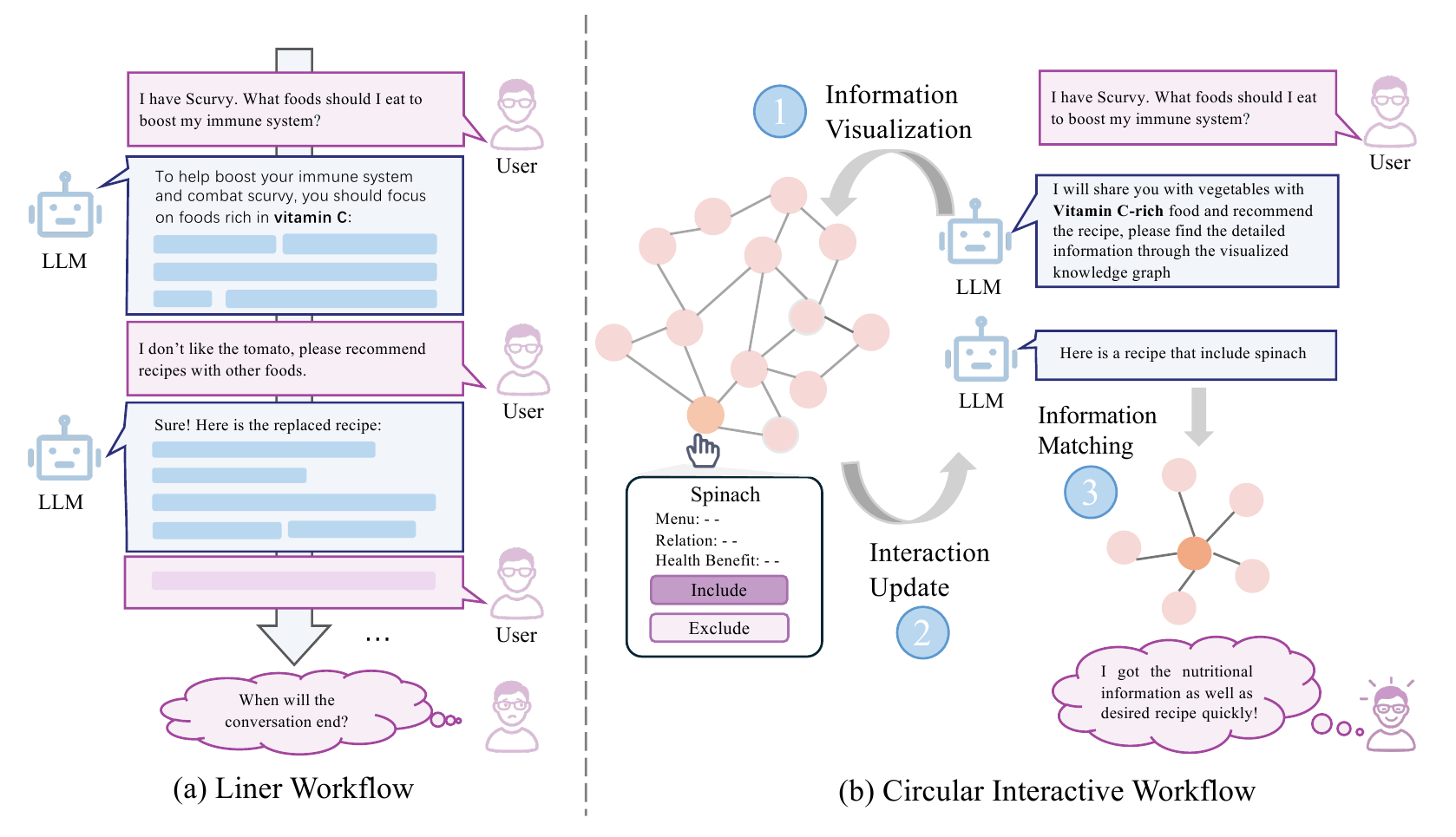}
    \caption{Unlike the traditional linear workflow (a) of LLMs, \systemName offers dietary guidance through interactive knowledge graph visualizations within a circular, interactive workflow (b), allowing users to quickly gain an overview of desired information.}
    \Description{Comparison of linear (a) and circular interactive workflows (b) in HealthGenie. In (a), a traditional linear interaction shows a user asking for dietary advice, receiving a response, and requesting a replacement for disliked ingredients, ending with uncertainty about the conversation's end. In (b), the circular workflow visualizes information through a knowledge graph, enabling users to interact with nodes like spinach, include or exclude ingredients, and match desired recipes efficiently, providing an overview of nutritional information and recipes in real-time.}
    \label{fig:teaser}
\end{teaserfigure}

%% file: section/1_introduction.tex
\section{Introduction}







Making informed dietary choices plays a crucial role in managing personal health and preventing chronic diseases. To make dietary information accessible and interpretable, research has focused on structuring this complex knowledge in an organized form --- \textbf{K}nowledge \textbf{G}raphs (KGs), which connect entities such as foods, nutrients, and health conditions through semantically meaningful relationships~\cite{he2019aloha, cui2023survey, abu2023healthcare, haussmann2019foodkg}. At the same time, the emergence of \textbf{L}arge \textbf{L}anguage \textbf{M}odels (LLMs) has revolutionized information-seeking behavior by offering interactive, conversational interfaces~\cite{yildirim2024multimodal, qiu2024llm, rajashekar2024human}. The combination of LLMs with KGs enhances both the usability of structured dietary knowledge and enables more intuitive, intelligent exploration of complex nutritional relationships~\cite{li2024preliminary,huang2023let}.

While integrating KGs can greatly enhance the quality of LLMs' outputs greatly, this benefit is underutilized with current text-based interfaces, which usually rely on linear text formats~\cite{chang2002effect}, such as extended paragraphs, and do not take advantage of the inherent structure of knowledge graphs by providing (i) visualized output~\cite{li2023knowledge, li2021kg4vis} and (ii) direct interaction~\cite{cox2020visualization, jiang2023graphologue, shah1999graphs}. First, existing LLM-based methods are capable of providing more detailed information through knowledge graphs~\cite{xu2024generate, wang2025knowledge, welz2024enhancing}. However, these text-based outputs are often too verbose, preventing users from easily focusing on the information that interests them and increases the time and effort required to identify desired content. For example, comparing the purposes of suggested dishes in a textual context is challenging and calls for visualizing relationships and unique features of each dish to assist users in making decisions. Second, in terms of direct interaction, users often need to engage in multiple rounds of conversation to eventually obtain their preferred recipes because text-based exchanges require them to independently determine their needs and input lengthy descriptions, which is time-consuming and inefficient. Therefore, developing a new interactive interface that facilitates decision-making and offers an intuitive method is critical. Combining KGs as the interface may provide a viable solution for dietary recommendations~\cite{yan2024knownet, oelen2024leveraging, xu2025interactive}. Such a combination can foster the advantages of each—leveraging the natural language capabilities of LLMs while retaining the structural clarity and expressiveness of KGs. 
In Figure~\ref{fig:teaser}, we illustrate an example comparing the conventional linear text-based approach (a) with our circular, knowledge-graph-driven interactive workflow (b). The left side shows how users typically must go through multiple rounds of text-based queries for dietary recommendations, while the right side demonstrates how visualizing and directly interacting with the knowledge graph can help them more efficiently refine or exclude certain ingredients in real time.

In this paper, we propose \systemName \xspace --- a novel interactive system that synergizes LLM-powered conversational interfaces with recipe-specific knowledge graph visualizations. \systemName is designed to empower everyday users by providing clear, evidence-based dietary recommendations. Drawing on insights from a formative study involving seven participants, our work addresses critical needs such as information transparency, reduced cognitive load, intuitive interaction, and personalized guidance. \systemName utilizes a circular workflow to enhance the LLM-KG-User interaction. Through this approach, the LLM identifies individual user preferences based on their interactions with the KG, dynamically incorporating these insights into its recommendations, which are then visualized back into the KG. By grounding LLM responses in a curated nutritional knowledge graph and actively offering intuitive visual guidance, \systemName enhances both the credibility and usability of dietary recommendations, ultimately supporting more informed recipe choices.

To evaluate how effectively \systemName facilitates knowledge graph visualization and intuitive interaction with LLMs for personalized dietary recommendation tasks, we conducted a within-subject evaluation with 12 experienced LLM users. We demonstrate the functionality of \systemName in four scenarios and guide users to complete the related tasks with our system and baseline system in a counterbalanced order. Both quantitative and qualitative results demonstrate that our system effectively delivers well-organized recipe information through integrated textual and visual representations while actively supporting users' preference-based decision-making. 

The main contributions of this work can be summarized as follows:
\begin{itemize}
    \item A formative study (N=7) that summarizes practices, challenges, and expectations on dietary information display. 
    \item \systemName, an interactive system featuring visualized KG response and intuitive interaction, empowering users in personalized dietary exploration.  
    \item An empirical user study (N=12) on the usefulness and effectiveness of \systemName in providing recipe recommendations and a discussion of insights derived from the user study for future LLM-KG integrated interface design.
\end{itemize}

%% file: section/2_related_work.tex
\section{Background \& Related Work}
In this section, we review prior work on the role of Artificial Intelligence (AI) in healthcare support systems, the current state of LLM-based conversational interfaces, the design of knowledge graph visualization and interaction—particularly within the Human-Computer Interaction (HCI) research community—as well as the limitations and challenges of LLM-driven nutrition and dietary guidance.

\subsection{AI-powered Healthcare Support Systems}
Traditional healthcare support systems focus on designing and implementing interactive technologies that provide users with the information they need to make informed decisions and manage their well-being~\cite{zhang2023understanding}. Early work shows that users commonly rely on search engines (e.g., Google and Yahoo) or social media platforms such as Quora and Reddit to obtain health information~\cite{jia2021online, milton2024seeking}. 
To address the need for professional healthcare consultation, more advanced AI tools have been explored. For example, Joshi et al. \cite{joshi2014supporting} developed an interactive voice response system to support individuals living with AIDS by providing healthcare tips. Similarly, Zhang \cite{zhang2023understanding} focused on the design of AI-driven interfaces to assist users in accessing healthcare information and consultations.


In recent years, large language models (LLMs) with task-agnostic architectures and extensive pre-training have seen wide application in both healthcare and nutrition \cite{turhan2024recipe}. Studies have suggested the potential of LLMs for providing informed decision support throughout clinical care, from diagnosis to treatment recommendations~\cite{kao2023assessing, rao2023evaluating, rao2023assessing}. Moreover, researchers have explored utilizing LLMs' superior natural language understanding and generation ability to engage in open-ended conversations with access to healthcare information~\cite{liao2024revolutionary, rajashekar2024human}. Despite the success of LLMs, the generative nature makes it unavoidable to result in hallucination, especially in the healthcare domain. To tackle this, Sachdeva et al.~\cite{sachdeva2024learnings} developed \textit{Build Your Own expert Bot platform} to create LLM-powered chatbots with integrated expert verification. These systems aim to provide more and more accurate information for non-expert users.

\subsection{Large Language Model-Based Conversational Interface}
In response to the rapid advancement of LLMs, the Human-Computer-Interaction (HCI) community has been actively 
exploring innovative conversational interface design to improve user interaction with LLMs across diverse applications~\cite{fan2024lessonplanner, yen2024memolet, chiang2024enhancing}. These efforts can be broadly categorized into two key research directions: (1) enhancing user comprehension of LLM-generated content and (2) leveraging LLM's open-ended conversational capabilities to provide task-specific guidance.

A prominent research focus involves designing systems that help users better understand and interpret LLM outputs. For example, \textit{Graphlogue}~\cite{jiang2023graphologue} converts text-based responses from LLM into graphical diagrams, facilitating information search and question-answering tasks. Similarly, \textit{WatiGPT}~\cite{xie2024waitgpt} proposed to empower users with enhanced comprehension and augmented control over analysis conducted by LLMs. \textit{FathomGPT}~\cite{khanal2024fathomgpt} supports interactive exploration of ocean data through intuitive visualizations. These systems aim to reduce cognitive load and improve transparency in human-LLM interactions.

Another key trend involves leveraging LLM's conversational flexibility to offer tailored assistance in specialized tasks. \textit{DiaryMate}~\cite{kim2024diarymate} assists users with reflective journaling by providing LLM-generated writing prompts and feedback. \textit{InkSync}~\cite{laban2024beyond} extends traditional chat-based conversational interfaces by integrating LLM-generated feedback during document editing, while \textit{Compeer}~\cite{liu2024compeer} introduces a proactive conversational agent that provides adaptive peer support. These systems show how LLMs can be harnessed to specific user tasks effectively.

Building upon these advancements, \systemName aligns with interactive conversational systems that enhance user understanding of LLM-generated responses. Aimed at reducing conversational turns, \systemName offers insights on how LLMs can support personalized dietary or recipe recommendations.

\subsection{Knowledge Graph Visualization and Interaction}

Fundamentally, a knowledge graph is a data model that represents knowledge as a set of nodes (i.e., entities), edges (i.e., relations between entities), and properties (i.e., attributes) that can be associated with both nodes and edges~\cite{agarwal-etal-2021-knowledge}. Due to their robust structure for reasoning over data, knowledge graphs have attracted growing interest in the research community, particularly in exploring how they can be leveraged to enable more effective visualizations and interactive interfaces.
Previous works have used KGs as common knowledge ground method and design visual interfaces to directly view and manipulate structured data for both experts~\cite{lemaignan2017artificial} and non-expert~\cite{zhang2021patterns,xia2021ktabulator, cai2024linking} users. 
For example, KGScope\cite{10418108} supports interactive visual exploration of knowledge graphs by providing embedding-based guidance to help users derive insights and navigate the broader network. Their evaluation demonstrates that s can offer valuable information and facilitate comprehensive exploration. Similarly, Ashby et al.\cite{ashby2023personalized} leverage both KGs and LLMs to create personalized, context-aware conversational experiences.

In professional domains such as healthcare - where precision and comprehensive analysis are critical --- KGs are especially valuable for offering hierarchical explanations of data. Santos et al.~\cite{santos2022knowledge} presents an open-source platform based on clinical knowledge graphs and provides extendable node edit functions to facilitate user demands. While Xu et al.~\cite{xu2025interactive} and You et al.~\cite{yan2024knownet} have successfully integrated KGs and LLMs for health information seeking but tend to overlook the importance of incorporating user interaction feedback and provide personalized recommendations.

\subsection{Large Language Models in Nutrition and Dietary Guidance}

Recent research has also explored the diverse applications of LLMs in nutrition-related tasks. Studies have demonstrated their potential in areas ranging from dietary data analysis to generating personalized nutritional recommendations for specific health conditions~\cite{sosa2024role, maida2024application}.
For example, some proposed applications envision LLMs as interactive tools that can answer nutrition-related questions in real-time, offering a readily accessible source of information for individuals seeking dietary advice~\cite{guo2025ai, kaloudis2025ai}. Moreover, \textit{ChatDiet}~\cite{yang2024chatdiet} employs LLMs to create tailored meal plan that consider a user's specific dietary restrictions, preferences, and underlying health conditions.

Despite recognizing the potential of LLMs, several studies have raised concerns regarding the accuracy and reliability of the information these models provide~\cite{niszczota2023credibility, bergling2025bytes}. 
The need for rigorous validation of LLM-generated nutritional advice by human experts has been widely emphasized to ensure the safety and efficacy of these tools. For instance, in collaboration with registered dietitians, Szymanski et al. \cite{szymanski2024integrating} developed The Food Product Nutrition Assistant, which generates detailed food product descriptions and personalized nutrition information. Another line of work explores the use of Retrieval-Augmented Generation (RAG) frameworks to improve LLMs' ability to interpret and apply established dietary guidelines \cite{dhaliwalnutribench}.

By establishing a professional corpus as its knowledge base, \systemName retrieve relevant and recommend healthy dietary with keyword extraction. It is designed to help users quickly access desired recipes with minimal mental effort by prioritizing reliable, transparent, and visually clear information compared to traditional long texts that often display disorganized columns that compromise readability, reliability, and transparency~\cite{milton2024seeking}.


%% file: section/3_formative_study.tex
\section{Formative Study}
We conducted a formative study to better understand users' needs for acquiring and exploring healthcare information and to identify the challenges they face when interacting with current conversational LLM interfaces. The insights gained from this study inform the design goals of our system.

\subsection{Setup}
\paragraph{Participants.} 
Seven participants (P1–P7) with diverse experiences in LLMs were recruited for the study. The group was comprised of two females and five males, aged between 25 and 34. Among them, three were casual users familiar with LLMs, and four were experienced users who employed LLMs daily to obtain information or make decisions. Additionally, concerning knowledge graphs, four participants had a strong understanding. In terms of health and diet management frequency:

\begin{itemize}
    \item 4 participants (P1-P4) actively manage their health and diet;
    \item 1 participant (P5) occasionally manage;
    \item 2 participants (P6-P7) hardly manage health and dietary.
\end{itemize}

\paragraph{Procedure.} To thoroughly understand users' challenges and demands in dietary recommendations, this formative study employs a two-stage procedure: a questionnaire and a co-design brainstorming session. First, participants completed a well-structured questionnaire about their experiences seeking nutrition information through search engines and LLMs, which took approximately 15 to 20 minutes. The aim was to identify the shortcomings of existing tools. Second, we presented our preliminary \systemName workflows and invited participants to co-design the system interface, encouraging them to share their opinions openly. In this phase, we sought to understand users' interaction requirements and expectations in real-world scenarios. All responses and discussions were recorded and transcribed for analysis.

\subsection{Findings}
Using the reflexive thematic analysis method \citep{clarke2017thematic}, we analyzed the recordings of each participant's viewpoints gathered from the questionnaire and co-design brainstorming session. All participants reported having experience using search engines (e.g., Google, Bing) to search for health information. Among them, six participants also used social media and conversational LLMs, with two participants occasionally consulting professionals such as doctors and nutritionists. Only one participant mentioned referring to books or academic papers. Here, we summarize three common types of information needs that emerged from the formative study.



\paragraph{\textbf{Visualized and Rapid Information Delivery.}} 
Participants expressed a strong preference for a system that not only presents information visually but also retrieves and displays results efficiently. Most participants preferred hybrid visualizations over text-only formats when seeking healthcare information. For example, P1 and P6 favored node graph diagrams due to their clarity in presenting concepts and logical relationships. P2 and P7 emphasized the usefulness of hierarchical visualizations, such as flashcards, for organizing complex content. While P5 preferred traditional text-only formats, P3 expressed concern that graphs alone might lack sufficient detail for full comprehension. Furthermore, several participants concurred on the necessity to minimize conversational turns and provide quick results. Participants P1, P2, P3, and P4 additionally desired features such as visual highlights to support rapid information catching, with P1 noting, ``Quick response is critical, as a user, I wish to obtain answers in a very short waiting time.''

\paragraph{\textbf{Trust and Transparency.}} 
Given the critical importance of healthcare consulting, the reliability of information sources is often a significant concern, especially those generated by LLMs. Participants P4, P5, and P7 highlighted that one of the biggest challenges when searching for healthcare information is distinguishing reliable sources from unreliable ones. They emphasized that the prevalence of misleading or conflicting information could complicate information exploration and undermine confidence and trust in LLM-driven tools. Furthermore, P2 mentioned that ``real-world data is hard to recognize,'' meaning that in the era of big data, recognizing and verifying real-world evidence remains a significant gap, particularly in domains where accuracy is paramount, which is also about the difficulties in distinguishing it from other types of information in a data-saturated environment. The application of KGs to present detailed relational information was suggested as a means to enhance the sense of trust and transparency.

\paragraph{\textbf{Personalized guidance.}} 
All participants anticipated an LLM-based interface designed for specific health information exploration, particularly one that offers inquiry generation and supports continuous interaction based on individual preferences. Specifically, P1, P3, P4, and P6 preferred a system capable of organizing conversations, predicting their intentions and interests, and suggesting inquiries to foster deeper discussions, thus going beyond merely answering questions. Conversely, P2, P5, and P7 emphasized the importance of maintaining the user's initiative, suggesting that the system should allow users to type their own questions while also providing selected query recommendations. P4 further suggested that ``the LLM agents should just deeply understand the user's context to address personal matters effectively.''


\subsection{Users' Preference on the Interface}
During the co-design brainstorming session, all participants expressed satisfaction with our design, which integrates a KG visualization alongside an LLM interaction interface. The KG effectively visualizes retrieval results, while the LLM-generated text responses outline key findings. P6 suggested adding a chat history feature to record users' query contents, enhancing recall and usability. Meanwhile, P3 recommended providing direct and tailored instructions to help users quickly understand the functions of the interface. P5 proposed making interactions with the KG nodes more dynamic, such as allowing users to rearrange or group healthcare information of interest for better organization. These suggestions contribute to a more well-rounded design direction for \systemName.

\subsection{Design Goals}
Based on insights from the formative study and participant feedback on \systemName's design, we establish the following \textbf{D}esign \textbf{G}oals (\textbf{DG}) to guide the development of our healthcare-focused, KG-enhanced conversational LLM with dynamic interaction capabilities.


\paragraph{\textbf{DG1. Support Information Visualizations: Both Graphs and Hierarchical Outlines}}
Non-expert users often struggle to comprehend nutritional information when it is presented in text format due to a lack of domain knowledge. Consequently, they may find it difficult to determine the next steps and may spend a considerable amount of time grasping basic concepts. To alleviate this issue, our system should incorporate KGs to visualize the relationships between complex concepts and employ structured outlines to present information in smaller, more manageable chunks~\cite{correia2025myaura}. This approach enables users to navigate from broad overviews to detailed specifics as needed. Such a strategy aligns with the design principle of \textit{progressive disclosure}~\cite{springer2019progressive}, which is often recommended in healthcare interfaces to manage complexity without obscuring important details. By using graphs and flashcards, users can more easily form an accurate mental model of the information structure instead of being confronted with an overwhelming wall of unstructured text.

\paragraph{\textbf{DG2. Provide Detailed Explanations with Professional Reasons.}} 
To enhance trust, LLM systems in healthcare must be designed with an emphasis on trust and transparency. A key principle is ensuring that the AI’s reasoning and explanations are accessible to users~\cite{larasati2021ai}. The system should deliver well-supported answers with logical and evidence-backed reasoning, enabling users to better understand the basis for its recommendations. Demonstrating explainability also addresses a common user concern: ``Where is this advice coming from?'' By structuring responses with clear justifications and contextual details, the system can alleviate doubts about accuracy and enhance overall reliability~\cite{procter2023holding}.

\paragraph{\textbf{DG3. Adapting Guidance to User Preferences.}} 
The development goal of our system is to assist users in navigating tailored recipes and dietary plans. To achieve this, guidance on personalized preference is critical, and prioritizing user preferences ensures the retrieval of highly relevant and personalized information. By deeply understanding user preferences (e.g., dietary restrictions, flavor preferences, or health goals), our system can refine search results, predict intentions more accurately, and deliver highly customized suggestions that align with individual needs~\cite{han2019deep, alvarado2022systematic, liu2023creative}.. Furthermore, 5 out of 7 participants emphasized the importance of AI systems that dynamically adapt recommendations to sustain engaging interactions. Leveraging intention prediction modeling, our system can proactively suggest relevant content such as recipe variations, nutritional insights, or meal-plan adjustments to maintain user interest. This approach ensures conversations remain contextually relevant and exploratory, encouraging users to engage more deeply with their dietary choices and overall health.



%% file: section/4_usage_senario.tex
\input{contents/02_workflow}

\section{Workflow and User Scenario} 
\subsection{Workflow}

Informed by the formative study and design goals, we design a \emph{circular interaction workflow} among the LLMs, KGs, and the user. In this workflow, the user initiates a query to the LLM, which retrieves relevant information from the KG, and the corresponding subgraphs are visualized. 
The user can then directly manipulate the visualized KG, such as by including or excluding specific entity nodes. This, in turn, affects the behavior of the LLMs, enabling a more adaptive and controllable interaction (see Figure~\ref{fig:workflow}).
This loop enables adaptive interactions by combining LLM reasoning with the visual, intuitive power of knowledge graphs. It allows users to explore AI suggestions while directly navigating and modifying structured information. The cyclical process supports dynamic, non-linear exploration, making it easier to access and refine details without a repetitive, linear approach. This \emph{circular interaction} (i.e., query $\rightarrow$ visualization $\rightarrow$ manipulation $\rightarrow$ refined query) synergizes the intelligent reasoning capabilities of LLMs with the intuitive, visual interactivity of knowledge graphs. Users are empowered to explore AI-generated suggestions while directly examining and modifying the underlying structured information space. The cyclical nature of this workflow also supports adaptive and non-linear explorations. Instead of relying on repetitive linear queries, users can iteratively refine information based on immediate visual feedback.

\input{contents/01_interface_overview}

Based on this workflow, we designed our system interface. As illustrated in Figure~\ref{fig:overall interface}, \systemName combines a chat panel (\textbf{A}) for conversational queries and responses with a query-generation panel (\textbf{B}) that helps users refine or rephrase their requests. The central area (\textbf{C}) displays a dynamic knowledge graph, highlighting entities such as recipes, ingredients, and nutritional benefits. Users can also control the granularity of the displayed subgraph via an adjustable slider (\textbf{D}), choosing whether to see ``show more'' or ``show less'' detail. Finally, the interaction record panel (\textbf{E}) keeps track of user actions (like including or excluding certain ingredients), which can be reviewed or undone at any time.

\subsection{Usage Scenario}

Below, we showcase how \systemName assists Caroline, an office worker who has been advised to reduce her protein and salt intake.

\paragraph{Initial Query and Response.}
Caroline begins by asking the system: \emph{``I plan to reduce protein and salt intake, please recommend some related recipes.''} As shown in Figure~\ref{fig:overall interface} (A), this query is processed by the LLM, which synthesizes an informative response explaining why low-protein and low-salt diets can alleviate the burden on the kidneys. Meanwhile, \systemName retrieves relevant items (recipes, ingredients, and nutritional facts) from the underlying knowledge graph (\textbf{DG1}).

\paragraph{Knowledge Graph Visualization.}
The interface then visualizes a subgraph of related ingredients and recipe nodes in the central panel (C), helping Caroline see potential meal options at a glance. She can expand or contract this network by adjusting the slider (D) to reveal more or fewer connected entities. A relevant explanation of the recommended recipe and ingredients will also be provided as part of the LLM's output (A), offering insights into their health benefits to enhance nutritional understanding and support better decision-making (\textbf{DG2}).

\paragraph{Refining Through Inclusion and Exclusion.}
While reviewing the suggested recipes, Caroline discovers that some recommendations still contain ingredients she prefers to avoid. As shown in Figure~\ref{fig:interactive}, she clicks the ``Black Pepper'' node to exclude it from further suggestions. \systemName then prompts her to \emph{Include} or \emph{Exclude} this item. Once she excludes ``Black Pepper,'' the action immediately appears in the interaction record panel (E). Similarly, she might include ``Crushed Tomato'' if she particularly wants to incorporate tomato-based dishes. After confirming these choices by clicking \emph{Apply}, the system updates the knowledge graph and regenerates refined recipe suggestions via the LLM, incorporating all of sher stated preferences.

\paragraph{Conversational Queries and Query Generation.}
Besides explicit inclusion/exclusion actions, Caroline can also ask open-ended questions (e.g., ``What are the nutritional values of these recipes?'') at any time, and \systemName will switch back to a conversational mode. It references both the current chat context and her interaction history to provide more detailed, personalized nutritional analyses. For more advanced or exploratory needs, Caroline may use the \emph{Query Generation} button (B) to let the LLM automatically propose queries based on her current dietary goals and interaction log, sparing her from having to articulate every question herself (\textbf{DG3)}.

\paragraph{Outcome.}
Through this circular interaction workflow—comprising initial queries, visualization, direct graph manipulation, and refined responses—Caroline arrives at a set of low-protein, low-salt recipes tailored to her preferences. She can dynamically expand or narrow down the recipe space, gaining a comprehensive understanding of each dish's key ingredients and potential health benefits. This intuitive, iterative process enables Caroline to make confident, well-informed dietary decisions that align with her medical advice and personal taste.

\input{contents/06_interactive_interface}

%% file: contents/02_workflow.tex
\begin{figure}
    \centering
    \includegraphics[width=\linewidth]{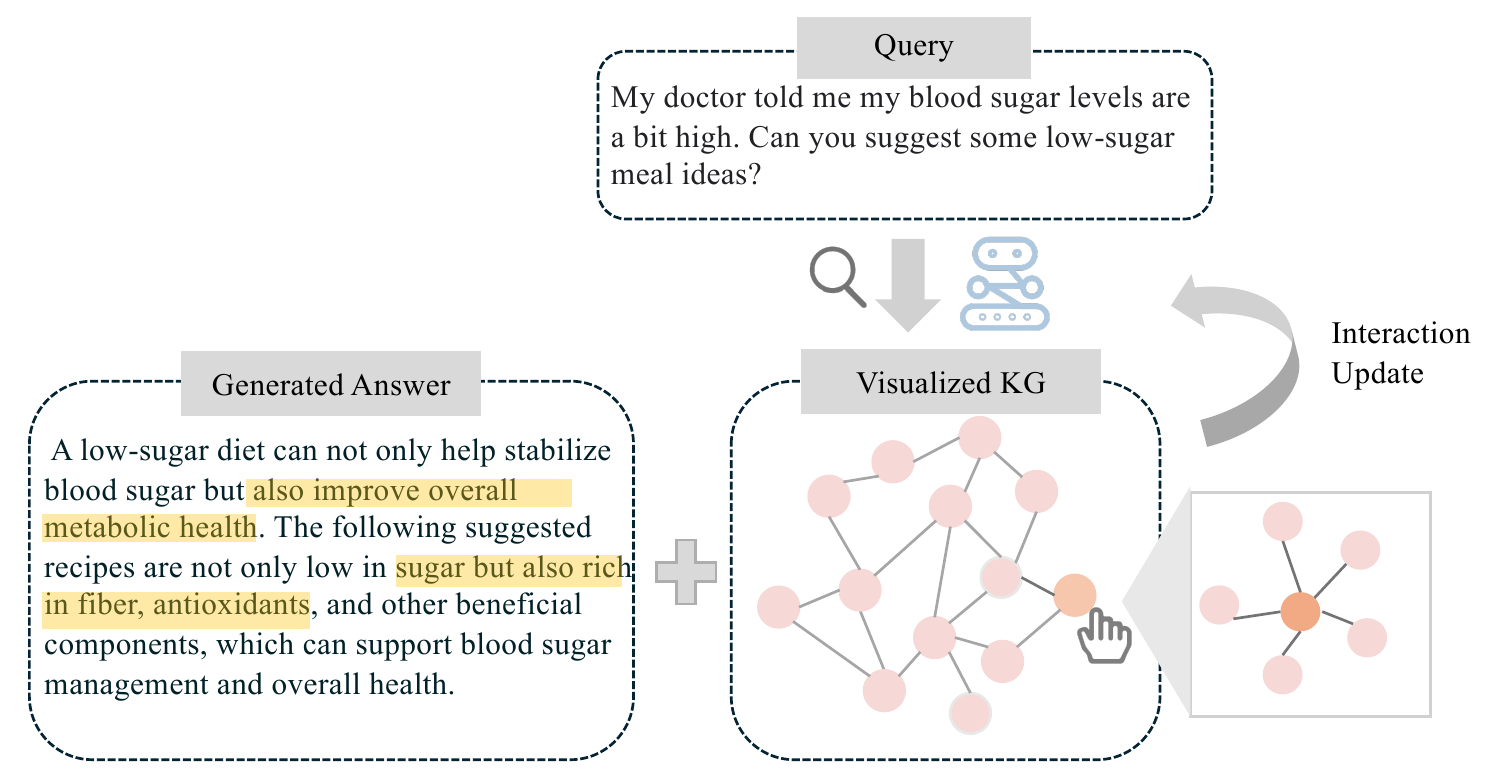}
    \caption{Circular Interaction Workflow: Users query an LLM, which retrieves and visualizes relevant knowledge graph (subgraphs); users can then manipulate the graph to iteratively refine outputs. Demonstrated via HealthGenie, this cyclical loop enables adaptive, non-linear exploration of dietary recommendations.}
    \Description{Circular interaction workflow for dietary recommendations. A user queries the LLM for low-sugar meal ideas to manage blood sugar levels. The system generates an answer emphasizing low-sugar diets' benefits for blood sugar stabilization and overall metabolic health. The relevant knowledge graph (KG) is visualized, and the user interacts with it by refining and exploring the options further. This cyclical, adaptive process enables non-linear exploration of personalized dietary recommendations via HealthGenie.}
    \label{fig:workflow}
\end{figure}

%% file: contents/01_interface_overview.tex
\begin{figure*}
\centering
\includegraphics[width=0.99\linewidth]
{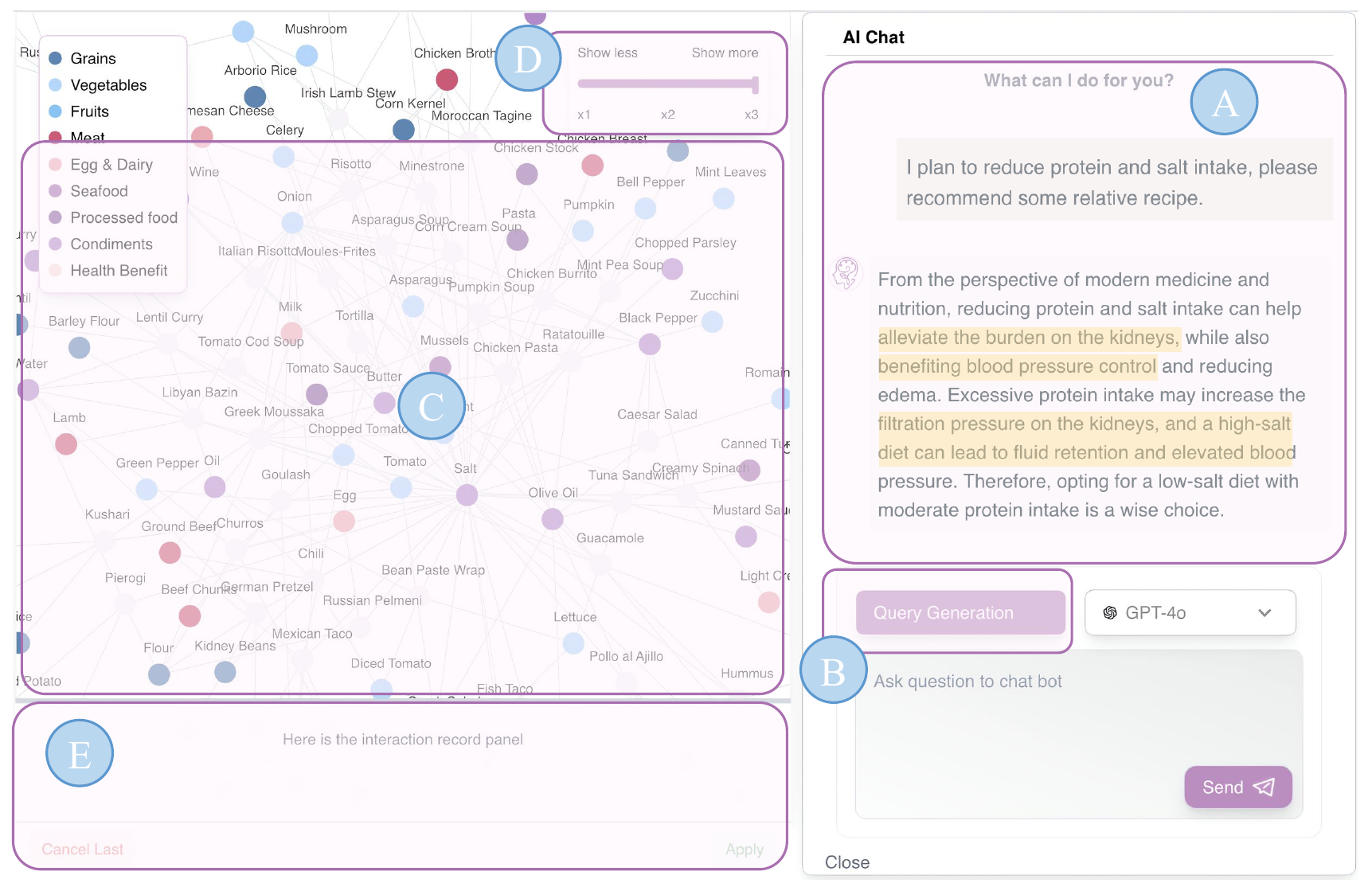}
\caption{The overview of HealthGenie interface, which integrates a visualized nutritional knowledge graph and a conversational dialogue system powered by an LLM. Users can initiate interactions by asking nutrition-related questions and requesting dietary guidance, then LLM retrieves relevant information and generates informative responses (A). Users can explore more relevant information using query generation (B). Simultaneously, the corresponding nutritional information is visualized within a dynamic knowledge graph (C), allowing users to explore with more and less information (D). \systemName provides interaction tracing visualization, supporting users to perceive and operate their deletion or addition intuitively (E).}
\Description{Overview of the HealthGenie interface. The user initiates a query (A) to reduce protein and salt intake, and the LLM generates a detailed response. Simultaneously, relevant nutritional information is displayed on a dynamic knowledge graph (C), which users can explore by showing or hiding elements (D). The interface also includes a query generation panel (B) for submitting questions, and an interaction record panel (E) allows users to cancel or modify actions. HealthGenie supports intuitive navigation of dietary recommendations through interactive visualizations.}
\label{fig:overall interface}
\end{figure*}

%% file: contents/06_interactive_interface.tex
\begin{figure}
    \centering
    \includegraphics[width=\linewidth]
    {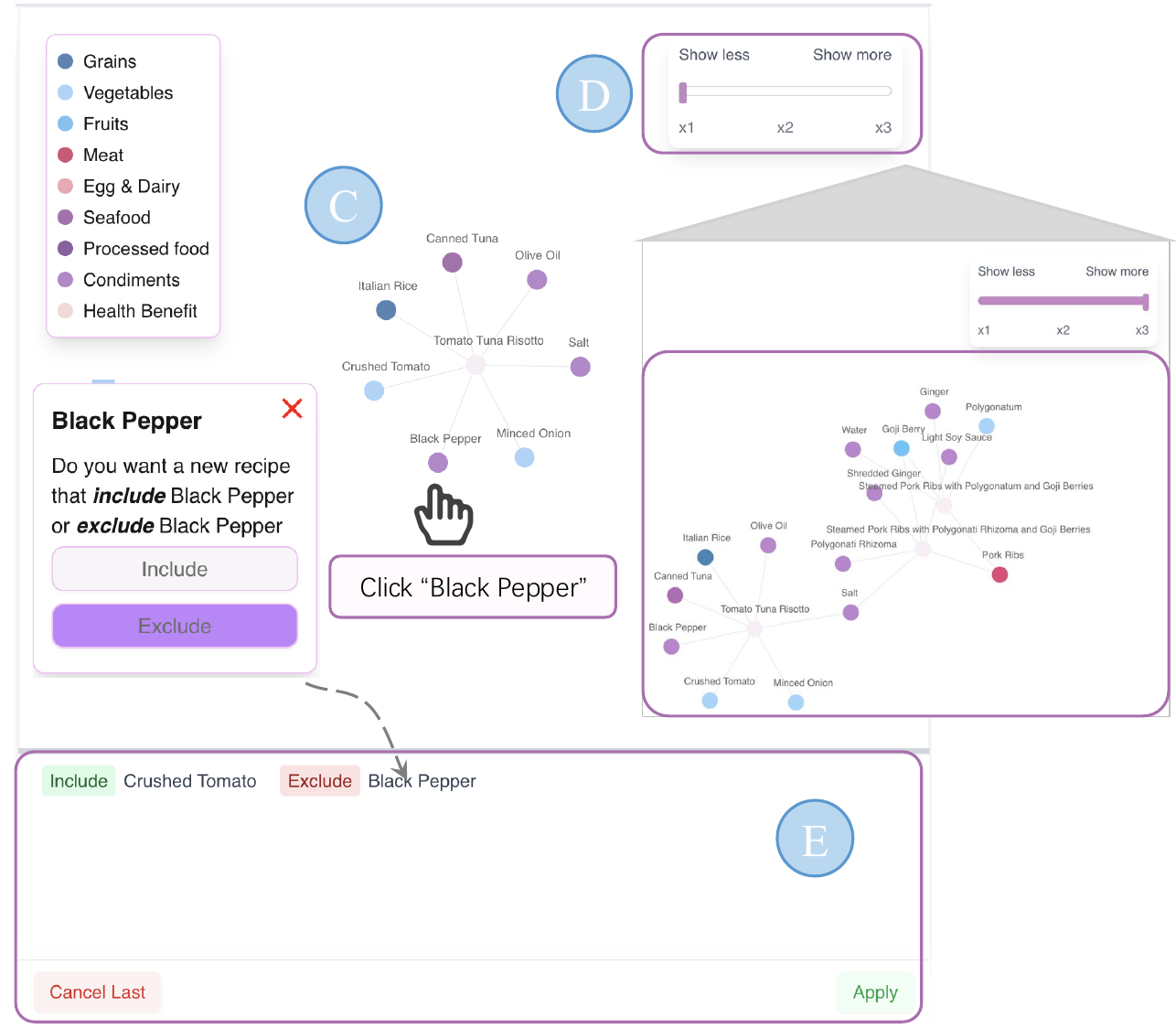}
    \caption{Interaction with the Knowledge Graph: The visualized Knowledge Graph in \systemName enables users to explore a broader range of recipes (D). Users can interact directly with the visualized nodes, such as hovering to view detailed relationships or clicking to include or exclude ingredients (C) in the next recommendation. Any selection to include or exclude ingredients is recorded in the interaction history panel (E), ensuring all user actions are tracked.}
    \Description{Interaction with the Knowledge Graph in HealthGenie. Users explore a broader range of recipes by interacting with the visualized knowledge graph (D). They can hover over nodes to view relationships or click on ingredients like Black Pepper to include or exclude them from future recipe recommendations (C). Any selections made are tracked and recorded in the interaction history panel (E), ensuring all user actions are documented.}
    \label{fig:interactive}
\end{figure}

%% file: section/5_system_design.tex
\section{Design and Implementation}
In this section, we present the design and implementation of \systemName, a sophisticated health recipe recommendation system that features a responsive user interface with real-time chat capabilities and interactive visualization of recipe knowledge graphs. \systemName is designed as a web application, where the front end is implemented using React and Next.js while the back end is created with Python and Flask. Specifically, we equip \systemName with multiple LLM configurations to provide answers: \textit{GPT-4o mini-2024-07-18}, \textit{DeepSeek-v3}, \textit{Claude-3.5-Haiku}, and \textit{LLaMA-3.2-90B}. Additionally, \systemName supports both English and Chinese.

The design of \systemName consists of four key components: Knowledge Graph Construction, User Query Processing, Knowledge Matching and Adaptive Output, and Interactive Feedback and Iteration. The system begins by constructing a recipe knowledge graph, which integrates a structured recipe database with ontologies covering ingredients, health benefits, and meal categories. When a user submits a query, the system analyzes the input, along with any available interaction history, to estimate the user's intent. Based on this intent, it retrieves a relevant subgraph and either generates a text-based response or simultaneously visualizes the subgraph while presenting personalized recipe recommendations. Users can interact with the visualized graph to refine their preferences, and these interactions are continuously fed back into the system to enhance future intent prediction and recommendation accuracy.

\subsection{Knowledge Graph Construction}
A meticulously developed knowledge graph underpins \systemName, ensuring that dietary recommendations remain transparent, explainable, and dynamically adaptable. We begin with a large-scale recipe repository containing approximately 12,500 unique recipes and 27,500 ingredient mentions. Each recipe, ingredient, nutrient, or dietary constraint is represented as a node, while edges capture semantic relations such as ``contains,'' ``belongsToCuisine,'' ``recommendsFor,'' or ``substitutableBy.'' In total, the KG now comprises over 100,000 nodes (up from the original 40,000) and at least 45 distinct relation types, enabling a comprehensive mapping of culinary and health-related concepts.

To enrich coverage, we apply a zero-shot LLM-based extraction approach that processes around 12 million tokens of free-text nutritional notes. This step identifies potential relationships beyond those explicitly defined in our ontology by leveraging the capacity of LLMs for synonym detection, context-aware inference, and domain-specific clustering. For example, if the LLM detects a statement such as ``lemon juice alleviates fishy odor,'' it can propose a new relation, ``neutralize Odor,'' which we then verify and integrate into the KG if deemed valid. By incorporating such inferred edges, the KG more accurately reflects nuanced dietary interactions that users may find relevant.

We store the KG in a hybrid format. A CSV-based index encodes each triple \textit{(subject, relation, object)}, enabling large-scale ingestion, while an in-memory representation supports high-speed pathfinding and subgraph extraction during user interactions. Nodes carry both numerical attributes (e.g., \texttt{calories=290}, \texttt{protein=12g}) and categorical labels (e.g., \texttt{allergenClass=shellfish}). This dual-layer architecture balances scalability and efficiency, which is essential for real-time recommendations. Versioning and incremental updates further enhance reliability: if users flag new allergies or preferences, \systemName revises the relevant edges or attributes, preserving an auditable history. Through iterative refinement driven by actual usage, the KG remains aligned with evolving nutritional best practices and personalized constraints.

\subsection{User Query Processing}
To interpret users’ natural language requests and map them to relevant graph entities, \systemName employs a domain-specific processing pipeline. Its primary objective is to capture a user's dietary goals, whether simple or complex, and translate them into symbolic constraints that can be matched against the KG. This design ensures that we can handle ambiguous or evolving requirements while upholding critical nutritional principles.

When a user initiates a query, \systemName orchestrates multiple steps to parse and contextualize intent, unify constraints, and deliver a precise system response. By synthesizing LLM outputs with symbolic parsing, the pipeline can robustly handle potential inconsistencies in user statements and respond with thoroughly verified recommendations. Whether the user needs a brand-new recipe suggestion or wishes to override a previously defined constraint, the query processing framework ensures that each demand is accurately documented and fed into subsequent matching and feedback routines.

\subsubsection{Intention Prediction}
Before performing detailed look-ups in the knowledge graph, the system first identifies the user's overarching intention using a specialized intent classification procedure. For instance, if a user requests ``Find me a vegan lunch under 400 kcal,'' \systemName determines whether this query represents a new recipe request, a modification of previously stated constraints, or a request for factual information. To achieve this, we employ a lightweight LLM-driven classification layer that distinguishes statements into \emph{recipe search}, \emph{constraint override}, \emph{information request}, or \emph{general clarification} categories.

Under the hood, \systemName utilizes a domain-specific lexical dictionary, enhanced with synonyms extracted through the LLM, to match phrases like ``tasty'' or `I like cheese'' to relevant symbols and to identify subjective or contradictory preferences. For example, if ``tasty'' indicates a user’s implicit desire for flavor-related attributes, the system prompts for clarification (e.g., ``Should we treat `tasty' as sweet, savory, or high in umami?''). By systematically mapping such discrete user demands to symbolic constraints, \systemName encodes high-level goals into parameters like \texttt{calorieCap=400} or \texttt{isVegan=true}. This approach enables the pipeline to manage potentially ambiguous or contradictory user requirements—such as simultaneously specifying ``vegan'' yet mentioning a liking for cheese—by formally acknowledging each constraint before deciding on conflict resolution strategies in subsequent steps.

\subsubsection{Query Parsing}
Once the system clarifies the user’s primary goal, it proceeds to a structured query parsing stage. Here, textual requests are converted into symbolic constraints that the KG can handle. Specifically, the parser identifies key entities (e.g., ``soy sauce,'' ``diabetes'') and relational modifiers (e.g., ``low in sodium,'' ``avoid dairy'') and then applies boundary values such as calorie or nutrient thresholds. By encoding user instructions in a formally defined constraint set, the system maintains consistency even when inputs are partially specified or ambiguous.

\paragraph{Keyword Extraction.}
In this phase, we rely on a domain-specific lexical resource and a lightweight morphological analyzer to extract entities from user queries. When an exact match is missing (for instance, the KG lacks an entry for ``broccolini''), \systemName consults an LLM-based synonym dictionary to propose a suitable substitute (e.g., ``broccoli''). Numerical expressions like ``under 300 calories'' or ``less than 10 grams of sugar'' are recognized and linked to the relevant attribute ranges in the KG. This approach ensures that each extracted term or numeric threshold is properly mapped to a node or edge attribute in the knowledge graph.

\paragraph{Language Processing.}
After extracting tokens, a shallow syntactic analysis interprets how the keywords relate to one another. For instance, ``I want more fiber'' indicates a positive inclusion constraint, whereas ``I dislike shrimp'' implies an exclusion. Additional instructions, such as ``cook them to retain nutrients,'' are translated into method-level flags (e.g., \texttt{highRetainNutrients=true}) that guide subsequent matching. The resulting constraint set $U$ supplies the system with a coherent, logically structured representation of user demands, ready for advanced filtering and recommendation steps in the knowledge graph pipeline.

\subsection{Knowledge Matching and Adaptive Output}
Having identified user constraints and recognized the relevant parts of the graph, \systemName proceeds to match requests against the KG, rank the results, and generate an appropriate textual or graphical response. Crucially, this output is not static. If the user later refines or reverses a constraint, the system must adapt the underlying logic in real time.

\subsubsection{Knowledge Matching}
The system first collects candidate nodes that appear likely to meet the user’s main criteria. For example, with $(\mathrm{vegan}=true, \mathrm{calorieCap}=400)$, it retrieves all dishes that contain no animal-derived nodes and have \texttt{calories} below 400. Edges such as ``containsIngredient'' or ``derivesFrom'' help confirm that each candidate truly aligns with the user’s constraints. Once the initial filter is complete, we refine the set by discarding nodes that violate any preference gleaned from earlier user interactions (e.g., a repeated dislike for tomatoes). In cases where multiple partial matches arise, such as dishes that meet nutritional needs but do not appear entirely vegan, \systemName explicitly labels them as ``borderline'' and can request user feedback. This mechanism ensures that subgraphs are computed not only by referencing static data but also by merging the user’s personal history to produce more tailored recommendations.

\subsubsection{Output Generation}
After identifying a matched subgraph, \systemName employs a multi-agent pipeline to synthesize a user-facing response. In particular, we integrate three key modules: (1) a \textit{Graph Retrieval Agent} that uses BFS-based subgraph extraction and symbolic checks to gather nodes conforming to user constraints, (2) a \textit{Relevance Scoring Agent} that ranks these nodes by nutritional and preference alignment, and (3) a \textit{Language Generation Agent} that composes a concise textual summary, leveraging LLMs for coherent phrasing. Specifically, the pipeline processes the subgraph to identify important dish properties, such as ``Grilled Tofu Wrap has 320 kcal, excludes dairy, and is moderately high in protein,'' and incorporates them into a concise explanation.

The system produces two synchronized outputs. First, a short textual summary clarifies why certain dishes were chosen, thus revealing how constraints like \texttt{calorieCap=400} or \texttt{isVegan=true} are satisfied. Second, an interactive graph visualization illustrates the relationships among recommended recipes, ingredients, and relevant health properties. Hovering on a node reveals attribute panels (e.g., ``low-sodium marinade''), and highlighting edges displays synergy or conflicts. This dual presentation, combining LLM-driven textual generation with targeted graph queries, helps users grasp both the overarching rationale and the specific interactions of food items.

\subsubsection{Adaptive Output}
Even after the system offers an initial recommendation, users often refine their constraints (e.g., ``remove soy sauce,'' ``add more greens''). In response, \systemName automatically revisits the relevant portion of the KG, revalidates previous selections, and applies newly discovered matches. This ``adaptive output'' cycle leverages adjacency-list lookups to filter nodes that conflict with updated constraints and to expand the search when new preferences suggest additional viable options. 
The system's real-time recalculation ensures minimal friction. For instance, when a conflict arises, such as an incompatible allergen or an overly restrictive calorie limit, visual nodes fade or disappear, and alternative edges or nodes are suggested. This iterative, feedback-driven workflow balances the creative flexibility provided by LLM-based text generation with the symbolic precision of knowledge-graph reasoning, ensuring that each final recommendation remains logically traceable to the user’s evolving preferences.

\subsection{Interactive Feedback and Iteration}
Because dietary choices are inherently personal and may shift as users explore various options, our system devotes special attention to fostering an iterative, user-centered feedback loop. This approach enriches the decision-making process by letting people constantly re-sculpt the set of recommendations according to taste, health conditions, or newly discovered constraints.

\subsubsection{Interactive Feedback}
To accommodate continuous dialogue and graph-based exploration, \systemName adopts a bidirectional feedback model.
If a user indicates hesitation through text, for example, ``I'm not sure if seitan is really gluten-free,'' the system can present clarifications or request confirmation, such as, ``Seitan typically contains wheat gluten; do you still want to include it?''
On the graphical side, hovering over or clicking on questionable items highlights exactly which properties might conflict with the user’s preferences. When the user chooses to remove or modify these items, the system logs that adjustment, updates the knowledge graph’s relevant edges or attributes, and reruns the matching routine. This interplay makes every user action interpretable and reversible. If the user regrets excluding a certain item, re-adding it is as simple as toggling a node back on the graph or removing the corresponding textual constraint.

\subsubsection{Recommendation Iteration}
Finally, after refining a preliminary recommendation, a user often wishes to broaden or narrow the search further—for example, exploring additional dishes that share some nutritional profile or removing entire categories of food if they prove unsatisfactory. This iterative ``recommendation evolution'' unfolds seamlessly. The system merges each newly specified preference into a symbolic profile store, ensuring subsequent queries or expansions reflect these updated conditions. Over time, \systemName learns a user's evolving preferences, such as frequently rejecting high-carb foods or consistently opting for leafy greens, and proactively suggests more relevant alternatives. This cyclical process culminates in a thoroughly personalized exploration of the culinary space. Users can experiment with a diversity of dishes while ensuring that each recommendation remains grounded in the knowledge graph and anchored in the constraints they have deliberately set.

%% file: section/6_evaluation.tex
\section{User Study}

We evaluated \systemName, deployed on a cloud server for participants to access, with 12 participants from various backgrounds. In this study, we focus on how effective and easy \systemName is to use in providing dietary recommendations under diverse health conditions and personal limitations. Specifically, we are interested in
the following research questions.

\begin{enumerate}
    \item \textbf{RQ1.} How do users perceive the visualized output and the explained texts from KG and LLM?
    \item \textbf{RQ2.} Can the proposed circle interactive workflow effectively support user preferences on recipe exploration?
    \item \textbf{RQ3.} How do users perceive the usefulness and the experience of interacting with \systemName?
\end{enumerate}

\subsection{Participants}
We recruited 12 participants (P1–P12) via social media, including 4 females and 8 males. Their ages ranged from 23 to 32 years (M = 27.4, SD = 3.02). Two participants held a bachelor's degree, four held a master's degree, and six held a doctorate. All participants had heard of knowledge graphs, and most reported understanding the general concept of combining LLMs with KGs (see Table~\ref{tab:participant_info}).

\begin{table}[ht]
\scriptsize
\centering
\begin{tabular}{cp{1cm}p{1.5cm}p{2cm}p{2cm}}
\hline
\textbf{ID} & \textbf{Education Level} & \textbf{Freq. of LLM Usage} & \textbf{Familiarity with KG} & \textbf{Familiarity with LLM-KG Integration} \\
\hline
P1 & PhD & Always  & Familiar & Familiar \\
P2 & PhD & Always & Unfamiliar & Unfamiliar \\
P3 & Bachelor & Often  & Know the concept & Know general idea \\
P4 & Master & Often & Familiar & Very Familiar \\
P5 & Master & Always  & Familiar & Very Familiar \\
P6 & Master & Always   & Familiar & Know general idea \\
P7 & Bachelor & Often  & Know the concept & Know general idea \\
P8 & PhD & Often & Familiar & Very Familiar \\
P9 & PhD & Always & Familiar & Very Familiar \\
P10 & PhD  & Often  & Know the concept& Know general idea \\
P11 & Master  & Always  & Know the concept & Know general idea \\
P12 & PhD & Always & Know the concept & Know general idea \\
\hline
\end{tabular}
\vspace{10pt}
\caption{Participant Information: We present participant ID, education level, LLM usage frequency and Familiarity with LLM-KG integration. }
\label{tab:participant_info}
\end{table}

\subsection{Study Design}
We carefully designed the study task to evaluate \systemName's usability regarding personalized dietary recommendations. The study includes four tasks across two different goals in total.

\paragraph{\textbf{Goal A: Personalized Dietary Recommendation}.} 
This task evaluates the system's ability to provide tailored dietary and recipe recommendations through interactive dialogue and knowledge graph visualization. Participants are instructed to engage with the system to achieve two objectives:

\begin{itemize}
    \item \textbf{Task 1} The participants are expected to identify a recipe that satisfies general nutritional requirements, adhere to dietary restrictions related to a health condition or dietary goal, and aligns with personal taste preferences through interaction.

    \textbf{Scenario Instruction.} \textit{Imagine you are under unhealthy conditions (e.g., indigestion, high blood sugar, poor sleep) or have some dietary goals (e.g., low-protein, low-salt, low-fat). Your objective is to obtain a dietary recommendation and recipe that not only satisfies general nutritional needs but also adheres to specific dietary restrictions associated with your condition. Additionally, you have personal food preferences that should be taken into account. You are encouraged to interact with the system until you receive a recipe that fulfills all three criteria: nutritional adequacy, medical suitability, and personal preference.}
    
    \item \textbf{Task 2} The participants aim to obtain a recipe recommendation derived from a predefined set of ingredients while also ensuring the recipe aligns with their personal preferences.

    \textbf{Scenario Instruction.} \textit{You are looking for a recipe recommendation based on a given set of ingredients you already have at home. In this goal, your primary aim is to obtain a recipe that utilizes the provided ingredients while aligning with your personal taste and health conditions. You should engage in the conversation and interaction with the system to refine its suggestions until a satisfactory recipe is found.}
\end{itemize}

\paragraph{\textbf{Goal B: Recipe Modification and Completion.}} 
This task examines the system's ability to support recipe adaptation and ingredient replacement in response to user needs or incomplete information with the following goals:

\begin{itemize}
    \item \textbf{Task 3} 
    Participants are tasked with modifying a recipe tailored to specific groups, ensuring it meets their dietary requirements while incorporating preferred ingredients.

    \textbf{Scenario Instruction.} \textit{You follow a vegan diet and have been provided with a non-vegan recipe. Your task is to adapt the original recipe so that it fully aligns with a vegan lifestyle. Additionally, you have specific ingredient preferences or constraints (e.g., allergies or dislikes). You should work with the system to modify the recipe accordingly, ensuring it respects both their dietary restrictions and ingredient preferences.}

    \item \textbf{Task 4} The participants are asked to complete an incomplete recipe by identifying and integrating suitable replacements for missing ingredients.

    \textbf{Scenario Instruction.} \textit{The provided recipe is incomplete. Your goal is to identify and request appropriate ingredient substitutions or obtain a revised version of the recipe that fills in the missing components. You should explore recommendations and engage in dialogue and interaction until a complete and satisfactory recipe is constructed.}
\end{itemize}

\subsection{Conditions}

\paragraph{\textbf{Baseline Condition.}} 
The baseline system consists of two components: a ChatGPT-based web application and a web interface with a knowledge graph identical to the one used in \systemName. For the conversational model, we employ GPT-4o. The knowledge graph interface offers basic retrieval functionality; for example, users can input an entity query and receive a subgraph related to the mentioned entity. In the baseline condition, participants are allowed to interact with both the ChatGPT application and the standalone knowledge graph interface to complete their tasks. They are free to decide when, whether, and how to use the provided tools.

\paragraph{\textbf{\systemName Condition.}} 
In the \systemName condition, participants are limited to using only the \systemName system to complete the assigned tasks. The conversational agent within \systemName is powered by the same GPT-4o model used in the baseline condition. Participants are required to engage with both the conversational agent and the interactive knowledge graph, which allows viewing and manipulating relevant subgraphs. As in the baseline condition, participants have the freedom to decide when and how to use the integrated system throughout the task.

\subsection{Procedure}
We conducted a within-subjects study to understand \systemName's effectiveness and user experience compared to the baseline system. The study begins with an introduction to \systemName and a description of its core functionalities. Then, the participants were asked to work on Task A (personalized dietary recommendation) and Task B (recipe modification and completion). To evaluate system effectiveness in a counterbalanced manner, participants are divided into two groups. In the first group, participants complete the tasks using \systemName first, followed by the baseline condition. In the second group, the order is reversed: participants begin with the baseline condition, then proceed to \systemName. After completing each task with a given system, participants are asked to fill out a well-designed, task-specific survey to provide feedback. Each task within a given condition, along with the corresponding survey, takes approximately 3–5 minutes to complete. Upon finishing all tasks across the different systems, participants are then asked to complete a questionnaire. This post-task survey focuses on participants’ overall preferences, satisfaction, and perceived effectiveness across \systemName, as well as on expressing their preferred features and challenges in using the system. Completion of this overall evaluation survey requires 35–45 minutes per participant.

\subsection{Measurements}
In the study, we measured users' experience both quantitatively and qualitatively using the following approaches.

\textbf{RQ1. Information Perception.} 
To investigate RQ1, we conducted a post-task evaluation study. First, users evaluated the quality of the sub-knowledge graph visualization and the generated text output across four dimensions: comprehensiveness, relevance, clarity, and organization. Next, users assessed the characteristics of the overall presented information in terms of accuracy, granularity, reliability, and interpretability. Finally, we asked users to rate the extent to which the visual information organization helped them accomplish each task.




\textbf{RQ2. User Preference Support.} 
To evaluate the effectiveness of \systemName’s personalized recommendation capabilities, we examined user satisfaction with personalized recommendations provided during the task. Participants were asked to rate how well the system’s suggestions matched their individual needs, preferences, and situational context. In the follow-up questionnaire, participants rated the system on three key dimensions \cite{alvarado2022systematic, knijnenburg2012explaining}: intuitiveness, personalization, and accuracy. Example statements included: \emph{``I can intuitively use the system to complete tasks''}, \emph{``The system's recommendations accurately align with my specific condition''}, and \emph{``The system's suggestions closely match my tastes and preferences.''} This metric serves as a critical indicator of how well \systemName supports user preference choice.




\textbf{RQ3. User Experience.} 
To address this question, we measured the task workload using the NASA Task Load Index~\cite{hart1988development} regarding \emph{Mental Demand}, \emph{Physical Demand}, \emph{Temporal Demand}, \emph{Performance}, and \emph{Frustration}. Additionally, we adopted the technology acceptance model~\cite{venkatesh2008technology} to assess perceived usefulness, ease of use, ease of learning, and enjoyment of interacting with \systemName, as well as participants' intention to use it. Participants also rated the perceived helpfulness of \systemName regarding dietary recommendations and recipe revision.

%% file: section/7_results_analysis.tex
\input{contents/05_matrix_bar}

\section{Results and Analysis}

In this section, we present both the quantitative and qualitative results corresponding to each research question (RQ). For each rating item, we first examine whether the order in which participants use the two interfaces influences the outcomes. To analyze RQ1 and RQ3, we employ a Wilcoxon signed-rank test~\cite{woolson2005wilcoxon} to compare the results between the two conditions. For RQ2, we conduct a series of mixed ANOVA tests~\cite{st1989analysis}, treating the order as a between-subjects variable and the interface type as a within-subjects variable. 

\subsection{RQ1. Information Perception}
Figure~\ref{fig:data_features} presents the characteristic results from the visualized KG and LLM outputs, highlighting the participants' responses to various evaluation dimensions. 
Overall, participants highly rated the output quality, especially in terms of \emph{Organization}. Both the KG and LLM outputs received high ratings for \emph{Organization} (KG: $M=4.66, SD=0.49, W=1.50, p=1.0$ and LLM: $M=4.67, SD=0.49, W=2.5, p=0.625$). This suggests that the presentation of information in a structured manner was highly effective across both systems.

Regarding data characteristics, \emph{Interpretability} received the highest ratings ($M=4.58, SD=0.51, W=2.00, p=1.00$), showing that participants found the outputs understandable and easy to follow. 
However, ratings for \emph{Accuracy} ($M=4.25, SD=0.75, W=9.00, p=0.1.00$) and \emph{Granularity} ($M=4.25, SD=0.62, W=2.71, p=0.042$) were slightly lower, with the latter indicating a statistically significant concern among participants regarding detail. This suggests that while the data were generally accurate and interpretable, there may be a preference for greater granularity in certain contexts.

Participants overwhelmingly agreed that the visual output substantially improved task efficiency. Notably, the data visualization received exceptionally high ratings for its effectiveness in facilitating task completion.
Participants strongly agreed that ``The visual output made revising the recipe easy'' ($M=4.91, SD=0.28$). Similarly, high ratings were given to statements that ``The visual output made understanding nutrition-related information easy'' and ``The visual output helped select preferred recipes quickly'' (both $M=4.88$), demonstrating the format's effectiveness for comprehension and decision-making.
Qualitative feedback further supported these findings. P4 emphasized, \emph{``It is easy to understand through the graph,''} confirming the value of visual aids. P7 noted, \emph{``I think the visualization is very beneficial during my interaction since it vividly presents the main content of the response with graphs,''} underscoring how graphical representation enhanced engagement. Additionally, P12 appreciated the detailed analysis, stating, \emph{``The analysis of my health condition and symptoms is remarkably thorough,''} suggesting participants valued both clarity and depth in the visual presentation.

\input{contents/03_combined_stacked_bar}

\subsection{RQ2. User Preference Support}

Figure~\ref{fig:baseline comparison} presents a comparative analysis of task completion ratings between \systemName and the baseline interface.
\systemName performed better than the baseline system in most of the tasks across different perspectives (Group 1 - \systemName: $M=4.51, SD=0.71,$, Baseline: $M=3.89, SD=0.76$; Group 2 - \systemName: $M=4.33, SD=0.98$, Baseline: $M=3.38, SD=1.29$). However, both systems demonstrated limited effectiveness in recipe recommendation tasks with constrained ingredients (\systemName: $M=2.58, SD=1.24$, Baseline: $M=2.833, SD=1.33$, $MD_{n}=-0.25, F=4.94, p=0.038$), in which users experienced frustration due to repeated ingredient clarifications before obtaining viable recommendations. This limitation may stem from challenges in accurately retrieving relevant recipes from the extensive knowledge graph, particularly when processing constrained ingredient inputs.



Notably, \systemName showed superior performance in accurately completing user tasks (\emph{Task Complete Accurately}), particularly for Task 2 (\systemName: $M=4.83, SD=0.38$; Baseline: $M=3.14, SD=1.98$; $MD_{n}=0.75, p=0.81$). This enhanced capability appears to derive from the system's user-centered design, which consistently incorporates individual health conditions and dietary goals into its recommendation algorithm. By maintaining persistent focus on these critical factors throughout multi-turn interactions, \systemName ensures nutritional requirements are both met and remembered across conversational contexts.



In personalized recommendation tasks, \systemName consistently outperformed the baseline system, with particularly strong results in Tasks 3 and 4 (\systemName:$M=4.67$, Baseline:$M=3.67$; $MD_{n}=1.00$). This performance advantage stems from \systemName’s cyclic workflow, which enables users to directly interact with and manipulate the recipe KG. This design not only fosters an engaging user experience but also allows the LLM to dynamically update and refine user preferences through iterative KG interactions, ensuring progressively better alignment with users’ evolving needs. Moreover, participants further reinforced these findings through positive feedback. For instance, P1 remarked: \emph{"I found it helps me to visualize the relationships of entities in a clear way; I would love to interact with the graph and explore interesting recipes."} Similarly, P2 highlighted the personalization of \systemName --- ``Personalizing desired or unwanted vegetables using knowledge graphs was really helpful to improve recipe.'' P8 also added that \emph{``I think the KG graph interaction feature is really impressive because it allows me to customize the content I want in a very straight way.''} These comments underscore the effectiveness of \systemName’s three-party interaction workflow (LLM-KG-User), which not only personalizes recommendations but also actively encourages user exploration and system engagement.

\input{contents/04_perceived}

\subsection{RQ3. User Experience}

Figure~\ref{fig:usefulness_perception} presents an overview of users' perceptions regarding the workload and interaction experience with \systemName. 
Users generally perceived moderate but comparable levels of effort across all demand dimensions, with no statistically significant differences in mental workload ($M=4.58, SD=2.02, W=10.00, p=1.00$), physical demands ($M=4.50, SD=1.88, W=5.0, p=0.62$), or temporal demands ($M=4.08, SD=1.24, W=0, p=1.00$). In contrast, participants reported high satisfaction with \systemName's performance ($M=5.67, SD=1.55, W=2.00, p=0.50$) and unanimously agreed that the system was easy to use for completing the assigned tasks.



Additionally, users consistently rated \systemName highly across key interaction experience dimensions. The system received good scores for Usefulness ($M=6.33, SD=0.49, W=1.50, p=1.00$), indicating unanimous agreement about its value in meeting user needs. Participants also reported exceptional Ease of Use ($M=6.4, SD=0.79, W=2.00, p=1.00$), confirming the interface's intuitive design. Enjoyment scores ($M=6.08, SD=0.90, W=3.00, p=0.18$) were notably high, suggesting users found the experience genuinely engaging. Finally, strong ratings for Orientation to Use ($M=6.0, SD=0.85, W=5.00, p=0.75$) demonstrated users' willingness to continue using \systemName, reflecting overall satisfaction with the system's usability and design.


\subsection{Challenges}

Despite the advanced usefulness and effectiveness of \systemName, from the open-ended experience sharing, we observed there exist three major challenges in using \systemName: (1) Insufficient data; (2) High latency; and (3) Limited KG structure.


\paragraph{Insufficient data.} Some participants reported that one of \systemName’s issues is that it occasionally provides unrelated information. For example, one participant stated, ``The range of recipes is somewhat limited and may lack knowledge of Eastern cuisines.'' This limitation is primarily due to our focus on supporting both Chinese and English systems. Initially, we prioritized Western recipes to ensure compatibility with both languages and translated the relevant entities into Chinese. However, translating Chinese recipes into English presents various challenges, such as multiple translations for the same ingredient. To maintain the quality and consistency of the recipes, we have primarily relied on Western culinary traditions.


\paragraph{High latency.} Other challenges associated with the performance of \systemName are reaction speed and latency. P8 mentioned that ``the latency is a little high,'' which we acknowledge as a valid concern. This issue arises because, in order to better detect user preferences and intentions, we have implemented multiple prompts within the system. This design allows us to gather more comprehensive and accurate feedback, helping us understand user behavior more clearly. However, to maintain consistency in the results and ensure that the system's output aligns with the user’s expectations, we deliberately chose this approach. It is important to note that this introduces a trade-off between speed and performance, as the system requires additional processing time to manage multiple prompts while still providing high-quality, relevant responses.


\paragraph{Limited KG structure.} Participants mentioned that they would prefer the KG to be organized by types or to suggest more related content. One participant even suggested, ``It would be even better if you could provide Amazon links to the products or recipe videos'' (P5). In our pre-built KG, recipes are currently organized by similar ingredients and health benefits, which we designed to simplify the information structure and facilitate quick retrieval. This approach was intended to ensure that users could easily find recipes based on their dietary needs or specific health goals.

%% file: contents/05_matrix_bar.tex
\begin{figure*}
    \centering
    \includegraphics[width=\linewidth]{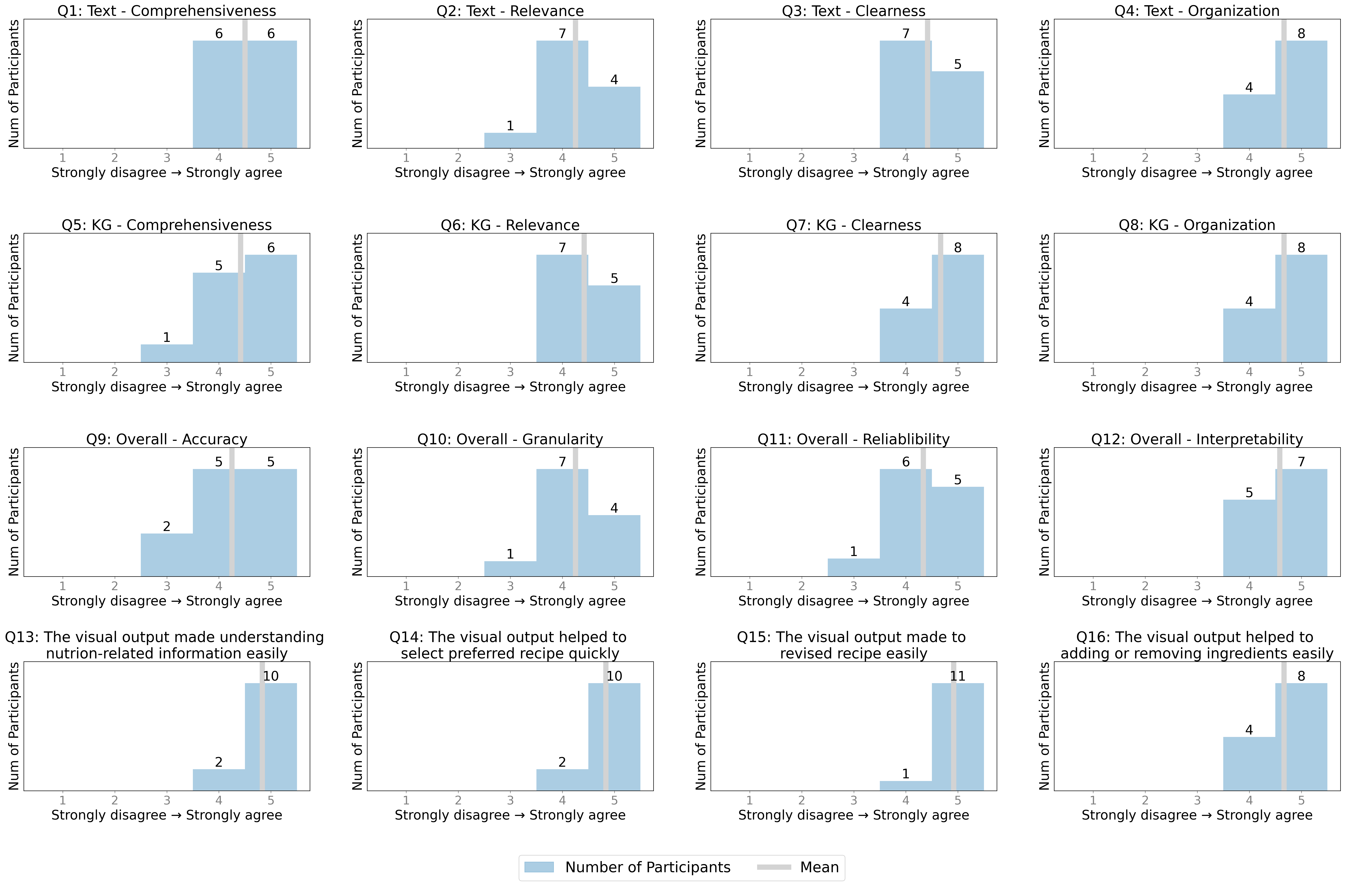}
    \caption{Participants' rating on output features of HealthGenie.}
    \Description{Participants' ratings on output features of HealthGenie. The bar charts display responses to various questions related to text and knowledge graph (KG) comprehensiveness, relevance, clarity, organization, and overall quality. Questions also assess the visual output's effectiveness in helping users understand nutrition-related information, select recipes, revise preferences, and add or remove ingredients. Ratings range from 'Strongly disagree' to 'Strongly agree', with the number of participants represented on the x-axis.}
    \label{fig:data_features}
\end{figure*}

%% file: contents/03_combined_stacked_bar.tex
\begin{figure}
    \centering
    \includegraphics[width=\linewidth]{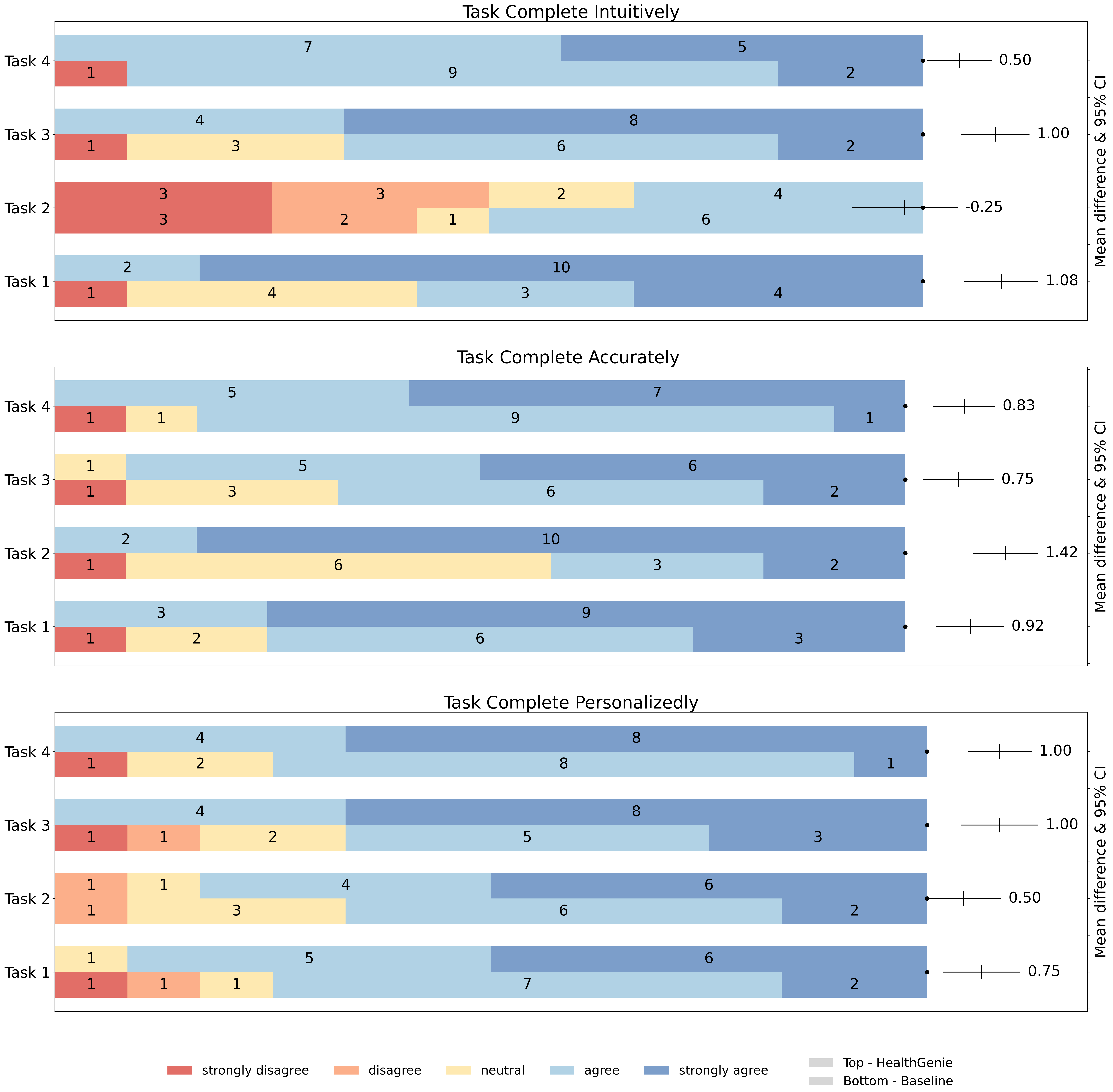}
    \caption{Participants' responses regarding the task completion intuition, task completion accuracy and task completion satisfaction among four tasks, measured by the 5-point Likert scale questionnaire for both based and our system. Bars present the mean differences of our system compared to the Baseline. Dots indicated the 95\% Confidence Interval.}
    \Description{Participants' responses regarding task completion intuition, accuracy, and satisfaction across four tasks. The horizontal bar charts display ratings from a 5-point Likert scale, comparing HealthGenie (Top) and Baseline (Bottom) systems. For each task, bars represent the frequency of responses, from 'strongly disagree' to 'strongly agree.' Mean differences between the systems are shown with 95\% confidence intervals, indicating how HealthGenie performed relative to the Baseline in terms of intuition, accuracy, and personalization.}
    \label{fig:baseline comparison}
\end{figure}

%% file: contents/04_perceived.tex
\begin{figure}
    \centering
    \includegraphics[width=\linewidth]{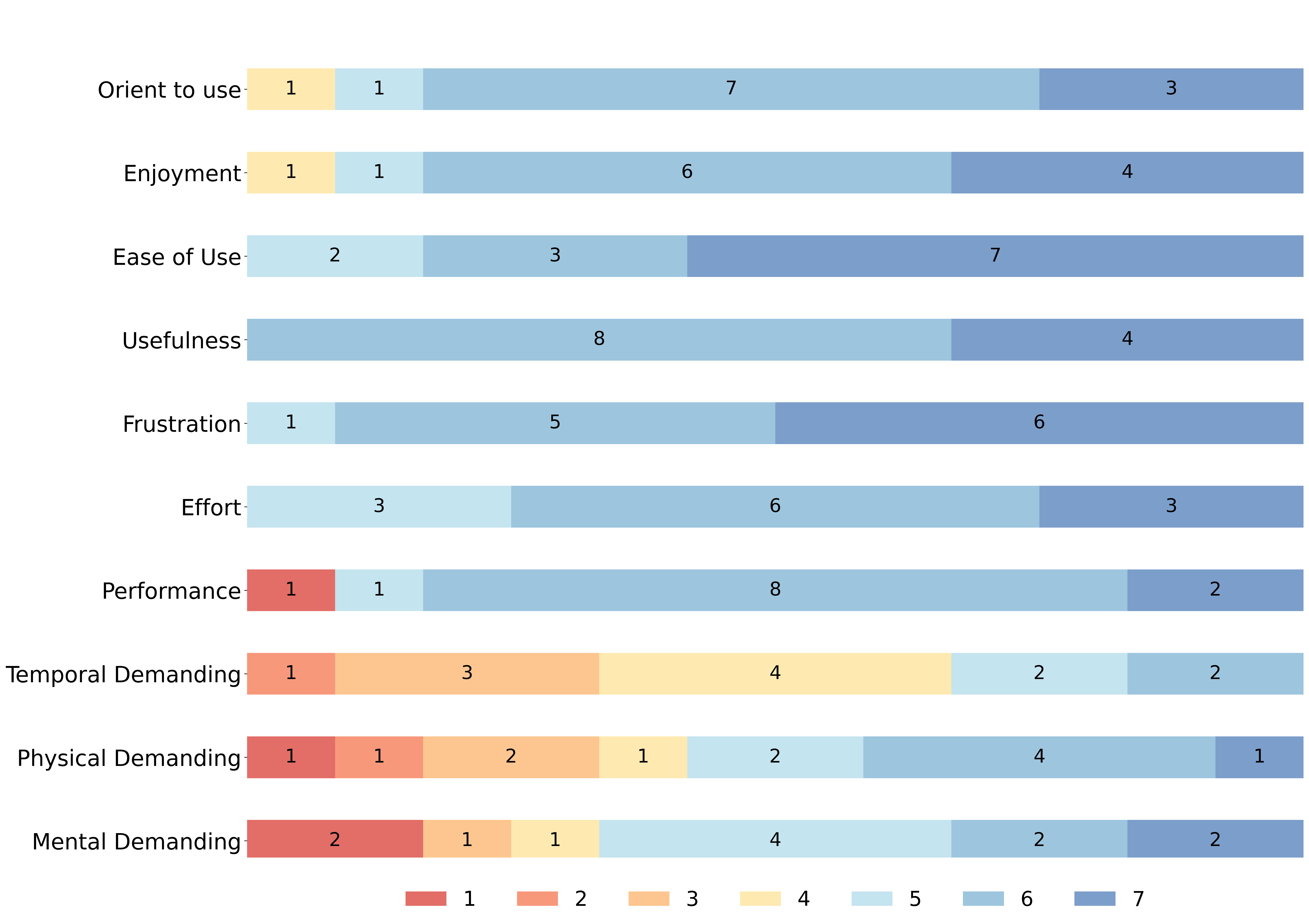}
    \caption{Participants' self-reported rating on usefulness and experience, measured by 7-point Likert scale.}
    \Description{Participants' self-reported ratings on usefulness and experience, measured by a 7-point Likert scale. The horizontal bar charts display ratings for various aspects of the system, including 'Ease of Use,' 'Usefulness,' 'Frustration,' 'Effort,' and 'Performance,' with responses ranging from '1' (strongly disagree) to '7' (strongly agree). Participants also rated temporal, physical, and mental demanding factors, with the frequency of responses represented in the bars.}
    \label{fig:usefulness_perception}
\end{figure}

%% file: section/8_discussion.tex
\section{Discussion}

In this work, we design and develop \systemName, an interactive system that provides personalized recipe recommendations and dietary guidance based on individual health conditions and goals, leveraging integrated KGs and LLMs. Our within-subjects study (N=12) compares \systemName with ChatGPT and a dummy KG retriever, thereby demonstrating its effectiveness in enhancing user support for meal planning decisions. Our user study reveals both the limitations and potential opportunities of \systemName, offering key design considerations for future interactive systems that integrate visualized KGs with intelligent LLMs.


\subsection{Visualizing Knowledge}
Visualization techniques are essential for enhancing the usability and interpretability of LLM-KG interfaces. In our work, we employ node-link diagrams~\cite{jiang2023graphologue, hearst2019would} to represent the structure and content of knowledge graphs, offering users an intuitive visual representation of entities and their relationships. 
However, effective KG visualization involves many additional considerations. Recent research has explored integrating KG and LLM reasoning~\cite{jiang2024kg, ji2024retrieval, tan2024paths, sun2023think}, where visualization can play a key role in presenting reasoning paths or evidence derived from the KG. This not only improves transparency but also helps users understand the system's decision-making process. 

On the other hand, \systemName’s visualization capabilities enable users to interactively explore the knowledge graph, fostering deeper insights. However, visualizing large-scale knowledge graphs remains challenging. KGs often contain millions of nodes and edges, making full-graph displays overwhelming and impractical. Future interface designs must address this complexity by incorporating techniques that simplify and focus visual representations, ensuring clarity without sacrificing information depth.


\subsection{Enriching User Interaction: Proactive Strategies and Beyond}

Effective LLM-KG interfaces should support diverse interaction paradigms that extend beyond basic question-answering. While our system currently enables interactive revision and exploration, user feedback highlights the need for richer engagement modes. A promising direction is proactive interaction, where the system anticipates user intent and delivers context-aware recommendations without explicit prompting. For example, in the health and biomedical domains, an LLM-KG interface could (1) propose relevant literature or diagnostic pathways based on a clinician’s ongoing case analysis~\cite{yan2024knownet}. (2) suggest personalized learning resources by tracking a student’s progress against a structured knowledge graph of medical concepts~\cite{o2024phenomics}. Such proactive features transform the interface from a passive tool into an intelligent collaborator, reducing cognitive load and accelerating decision-making. However, implementing these capabilities requires careful design to balance automation with user control. Future work can explore adaptive triggering mechanisms.



Effective LLM-KG interfaces must evolve beyond basic question-answering to support more sophisticated interaction paradigms. While our current system enables interactive exploration and revision, user feedback reveals opportunities for richer engagement through proactive interaction—where the system anticipates needs and delivers context-aware recommendations. 
For example, in the health and biomedical domains, an LLM-KG interface could: (1) propose relevant literature or diagnostic pathways based on a clinician’s ongoing case analysis~\cite{yan2024knownet}; and (2) suggest personalized learning resources by tracking a student’s progress against a structured knowledge graph of medical concepts~\cite{o2024phenomics}. 
Such capabilities would transform the interface from a passive tool to an active collaborator, reducing cognitive load while accelerating decision-making.

However, implementing proactive features requires careful design to balance automation with user control. The system must provide timely, relevant assistance without becoming intrusive or overbearing. Future work should explore adaptive triggering mechanisms that respect user preferences while maintaining transparency about the system's reasoning process. This includes developing robust intent prediction models and designing recommendation interfaces that preserve user agency—critical considerations for building trust in proactive LLM-KG systems.




\subsection{Personalized Knowledge Graph Construction}


While \systemName provides personalized recommendations through external knowledge sources, its effectiveness is inherently constrained by the static nature of pre-existing knowledge graphs, which cannot fully capture individual users' unique needs and contexts. To overcome this limitation, we propose empowering users to construct and refine their own knowledge graphs by incorporating personal data sources such as documents, notes, and domain-specific materials~\cite{zhao2024agentigraph,guo2024lightrag}. Future developments should focus on three key aspects: (1) intuitive interfaces for data ingestion and KG visualization, (2) LLM-assisted entity extraction and relationship identification from unstructured personal data, and (3) interactive tools for knowledge refinement and maintenance. This approach would enable users to not only build personalized knowledge structures but also actively curate and evolve them over time, creating dynamic representations that better reflect their individual knowledge domains and requirements.

%% file: section/9_conclusion.tex
\section{Conclusion}

We present \systemName, a novel interactive system that enhances dietary recommendations by dynamically integrating KGs with LLMs. Our approach enables users to visually explore nutrient relationships, filter ingredient options by health constraints, and refine preferences through direct graph interactions, thereby eliminating the need for verbose text exchanges. 
Our user study demonstrates that \systemName provides more intuitive and organized recommendations compared to traditional text-based interfaces, with high user satisfaction and perceived usefulness. By bridging expert-level dietary knowledge with everyday decision-making through structured, interactive visualization, this work advances the design of human-AI interfaces for personalized health guidance.